\documentclass[twocolumn,showpacs,floatfix,aps]{revtex4}

\usepackage{graphicx}
\usepackage{dcolumn}
\usepackage{bm}

\begin{document}
\title{Microwave-induced flow of vortices in long Josephson junctions}

\author{F.~L.~Barkov\footnote{Permanent address:
Institute of Solid State Physics, Chernogolovka, Moscow Region,
142432, Russia }, M.~V.~Fistul\footnote{Present address:
Theoretische Physik III, Ruhr-Universit\"{a}t Bochum, D-44801
Bochum, Germany} and A.~V.~Ustinov}
\address{Physikalisches Institut III, Universit\"{a}t
Erlangen-N\"{u}rnberg, D-91058, Erlangen, Germany}

\date{\today}

\begin{abstract}
We report experimental and numerical study of microwave-induced
flow of vortices in long Josephson junctions at zero dc magnetic
field. Our intriguing observation is that applying an ac-bias of a
small frequency $f \ll f_p $ and sufficiently large amplitude
changes the current-voltage characteristics ($I$-$V$ curve) of the
junction in a way similar to the effect of dc magnetic field, well
known as the flux-flow behavior. The characteristic voltage $V$ of
this low voltage branch increases with the power $P$ of microwave
radiation as $V_{s}\propto  P^{\alpha}$ with the index $\alpha
\simeq 0.5 $. Experiments using a low-temperature laser scanning
microscope unambiguously indicate the motion of Josephson vortices
driven by microwaves. Numerical simulations agree with the
experimental data and show strongly {\it irregular} vortex motion.
We explain our results by exploiting an analogy between the
microwave-induced vortex flow in long Josephson junctions and
incoherent multi-photon absorption in small Josephson junctions in
the presence of large thermal fluctuations. In the case of long
Josephson junctions the spatially-temporal chaos in the vortex
motion mimics the thermal fluctuations. In accordance with this
analogy, a control of the intensity of chaos in a long junction by
changing its damping constant leads to a pronounced change in the
shape of the $I$-$V$ curve. Our results provide a possible
explanation to previously measured but not yet understood
microwave-driven properties of intrinsic Josephson junctions in
high-temperature superconductors.
\end{abstract}


\maketitle

\section{Introduction}

The discovery of intrinsic Josephson junctions in highly
anisotropic high-$T_c$ superconductors~\cite{kleiner92} has
boosted the interest in the dynamics of Josephson vortices. These
junctions were proposed as candidates for high power flux-flow
oscillators at THz frequencies. The Josephson penetration depth
$\lambda_{\rm J}$ in these junctions is rather small, typically of
the order of $0.1\,\mu$m. Thus the standard micron-size intrinsic
junctions possess motion of Josephson vortices, similarly to
conventional low-$T_c$ long ($\ell\gg \lambda_J$) junctions.

Many of the experiments with intrinsic junctions have been
performed during the last decade at microwave (GHz) frequencies,
which are well below the junction's zero-bias plasma frequency
$f_{\rm p}$. These experiments led to observations of Josephson
plasma resonance \cite{plasma-res-HTS} and novel microwave-induced
branches in the current-voltage  characteristics ($I$-$V$ curves)
of intrinsic junctions \cite{HTS-rf-branches}. While the Josephson
plasma resonance has been thoroughly investigated and explained,
the effects produced by microwaves on $I$-$V$ curves have not yet
been fully understood. One obvious reason for that is that the
dynamics of Josephson junctions at frequencies $f<f_{\rm p}$ can
be chaotic and thus complicated. Nonetheless, many experiments
with underdamped intrinsic junctions driven by microwaves at zero
dc magnetic field revealed very puzzling {\it smooth branches} in
$I$-$V$ curves \cite{HTS-rf-branches}. The aim of our work is to
find an explanation for this behavior. For this purpose, we have
performed a detailed study of conventional low-$T_c$ long
Josephson junctions under strong microwave irradiation and have
done also numerical simulations, which we compare with
experimental data.

The dynamics of long Josephson junctions under microwave
irradiation was studied in various contexts in the past both
theoretically and experimentally. In particular, there had been an
extensive search for regimes in vortex motion synchronized by the
external microwaves \cite{Costabile}. The vortex dynamics in
underdamped junctions under the influence of ac drive is often
very complicated, showing along with synchronized (phase-locked)
dynamics \cite{locking} also the chaotic behavior
\cite{chaos-LJJ-with-RF}. The spatially-temporal chaos in
underdamped long Josephson junctions typically occurs at low drive
frequencies ($f<f_{\rm p}$). Moreover, the chaotic behavior in
long junctions may appear even in the absence of external
irradiation at low characteristic frequencies of vortex
oscillations \cite{chaos-LJJ-no-RF}. Increase of damping usually
helps obtaining more regular dynamics.

The interest to drive vortices in long junctions by microwaves has
been also biased by practical applications. Recently,
low-frequency microwave properties of long Josephson junctions
have been studied in relation to their possible application in
tuneable resonators \cite{Goldobin-2002}. On the other hand,
driving long junctions by microwaves is now successfully used for
phase-locking of Josephson flux-flow oscillators in integrated
superconducting sub-millimeter wave receivers
\cite{Kosh-Shitov-2000}.

In this paper, we investigate experimentally (Section II) and
numerically (Section III) the effect of microwave irradiation on
{\it long} underdamped Josephson junctions. In particular, we are
interested in the influence of microwaves on the current-voltage
characteristics of the junctions at zero (or small) dc magnetic
field. Our results suggest an explanation for previously observed
microwave-induced smooth branches in $I$-$V$ characteristics of
intrinsic high-$T_c$ junctions. We find that these smooth branches
are due to flux flow, but the vortex motion is very irregular and
shows spatially-temporal chaotic behavior.

It is interesting to note that the smooth branches in the $I$-$V$
curves of long junctions resemble the ones observed in extremely
small Josephson junctions driven by microwaves of a small
frequency in the presence of large thermal fluctuations
\cite{KovalFistUSt}. In this case the effect of microwaves on the
$I$-$V$ curves has been explained as incoherent multi-photon
absorption in the Josephson phase diffusion regime. Thus, we will
discuss (Section IV) our results by exploiting the analogy between
the spatially-temporal chaos of vortex motion in long junctions
and the diffusive behavior of the Josephson induced by thermal
fluctuations in small Josephson junctions.

\section{Experiments}

We performed measurements using Nb/Al-AlO$_x$/Nb long Josephson
junctions \cite{Jena} of overlap geometry. Below we present our
results for the junction of the length $L=250\,{\rm \mu m}$ and
width $w=3.5\,{\rm \mu m}$. The critical current density $j_c$ is
about $180\,{\rm A/cm^2}$, which corresponds to the Josephson
penetration length $\lambda_{\rm J}$ of about $28\,{\rm \mu m}$
and zero-bias plasma frequency of $f_{\rm p}\approx 77\,$GHz. The
sample was placed inside the vacuum volume of a
variable-temperature optical cryostat. Magnetic field was applied
in the plane of the junction perpendicular to its larger
dimension. The dc measurements were carried out using a four-probe
configuration.

We investigated the junction behavior under the microwave
irradiation of a frequency in the range from $10\,$GHz to
$20\,$GHz at two different temperatures $4.2\,$K and $5.0\,$K. The
external microwave radiation was applied to the junction by an
antenna placed outside the cryostat close to the optical window
facing the sample. Thus, the ac bias current was induced by the
antenna in the bias leads and across the junction. The typical
power $P$ of the applied microwave radiation was ranging from few
pW to several mW, as referenced to the output port of the
microwave source. Below we present and discuss the results
obtained for the frequency of microwave radiation
$f\equiv\omega/(2\pi)=16.61\,{\rm GHz}$, which corresponds to the
$f \simeq 0.2 f_{\rm p}$. We note that results obtained at
different frequencies within the studied range were qualitatively
similar.

\begin{figure}[tbp]
\begin{center}
\includegraphics[width=3.2in]{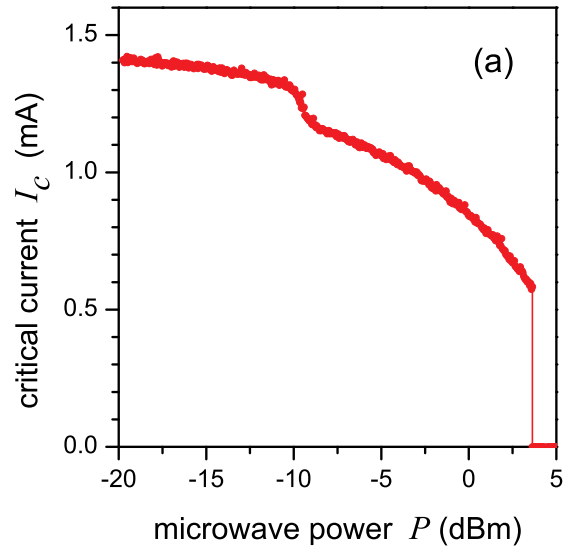}
\includegraphics[width=3.2in]{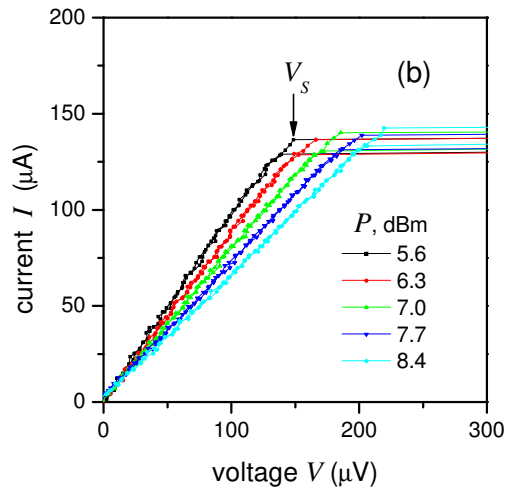}
\includegraphics[width=3.2in]{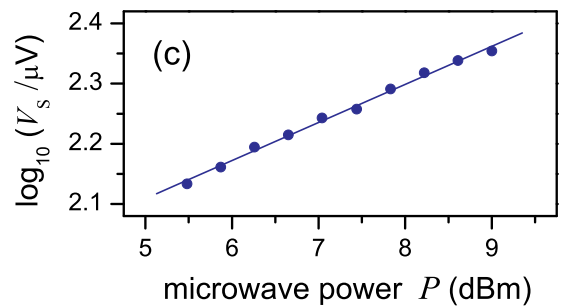}
\caption{(a) Critical current $I_c$ of the junction at zero
magnetic field versus the applied microwave power $P$ at the
frequency $f=16.61\,{\rm GHz}$. (b) $I$-$V$ curves for
different power levels at $P>P_{cr}$. The measured voltage $V_s$
at the top of the first curve is also shown. (c) Dependence of the
voltage $V_s$ on the microwave power $P$ using double-logarithmic
scale.} \label{Fig1}
\end{center}
\end{figure}

The dependence of the critical current of the junction on
microwave power at zero magnetic field and temperature $4.2\,{\rm
K}$ is shown in Fig.~\ref{Fig1}(a). The critical current $I_c$
monotonically decreases with power and drops abruptly to zero at
the power level $P_{cr}\approx 3\,{\rm dBm}$. This kind of
behavior can be interpreted in general terms for nonlinear systems
as a threshold in energy transmission \cite{Leon-2002}. For a long
Josephson junction, it can be treated as ac-driven dissociation of
a virtual breather (vortex-antivortex bound state) at the junction
boundary \cite{Borya-Misha}. A smoother drop occurs at the power
level of $P_{r}\approx -10\,{\rm dBm}$. It is associated with the
Josephson plasma resonance, which we will not address in the rest
of the paper. Plasma resonance in long junctions has been studied
previously in several experiments
\cite{plasma-LJJ,Pedersen-Sakai}.

As the applied microwave power exceeds $P_{cr}$, the supercurrent
branch of the $I$-$V$ curve disappears. Fig.~\ref{Fig1}(b) shows
the gradual changes in $I$-$V$ curves with increasing microwave
power above $P_{cr}$. Each $I$-$V$ curve has a linear region at
low bias and its slope decreases with power. At a certain bias
point, which depends on the power level, we observe a sudden jump
from the voltage $V=V_s$ to much higher voltages of the energy gap
at $V\approx 2.7\,$mV. The dependence of the voltage $V_s$ of this
critical point on the applied power is presented in
Fig.~\ref{Fig1}(c). We have found this dependence to be power-law
$V_s\propto P^{\alpha}$ with the index $\alpha=0.55\pm 0.01$. Very
similar behavior of the $I$-$V$ curves was also observed at the
temperature $5\,{\rm K}$. The major difference was in the modified
shape of the $I$-$V$ curves, where the linear branch at low bias
was replaced by a more pronounced pseudo-resonant feature, i.e. a
peak with downward curvature. Figure~\ref{Fig2}(a) illustrates
this behavior. The microwave power dependence of the switching
voltage $V_s$ (its position is marked in Fig.~\ref{Fig2}(a))
remained power-law with $\alpha=0.62\pm 0.01$.

It is worth mentioning that there is some uncertainty in
determining the position of the switching voltage $V_s$
(especially in the case of numerically calculated curves - see
results below). At the same time, it will be useful in the
following to have a criterion for the voltage which characterizes
the effect of microwaves on the $I$-$V$ curves. One could, for
example, choose some current value and observe how the voltage
$V_I$ at this current changes with microwave power $P$. In such a
case, however, question arises whether the behavior of $V_I(P)$ is
universal or depends on current. We have investigated $V_I(P)$
dependence for three current values marked by dotted horizontal
lines in Fig.~\ref{Fig2}(a). The dependence is always power-law
$V_I\propto P^{\beta}$ at $\beta=0.68$, $0.63$ and $0.47$ for
current values 150; 184 and 223 $\mu{\rm A}$ correspondingly.
Thus, the power-law $V\propto P^{x}$, where $V$ is either $V_s$ or
$V_I$ and value of $x$ is close to 0.5, describes the IV-curve
evolution versus applied microwave power with a reasonable
accuracy.

It is interesting that the $I$-$V$ curves under the influence of
microwave irradiation resemble the well known flux-flow state of
long Josephson junctions. The microwave field $~\sqrt{P}$ affects
the junction similarly to dc magnetic field $H$ applied in the
plane of the tunnel barrier. However, the sudden drop of the
junction critical current $I_c$ to zero at $P=P_{cr}$
qualitatively differs from behavior of $I_c$ in dc magnetic field.

\begin{figure}[tbp]
\begin{center}
\includegraphics[width=3.2in]{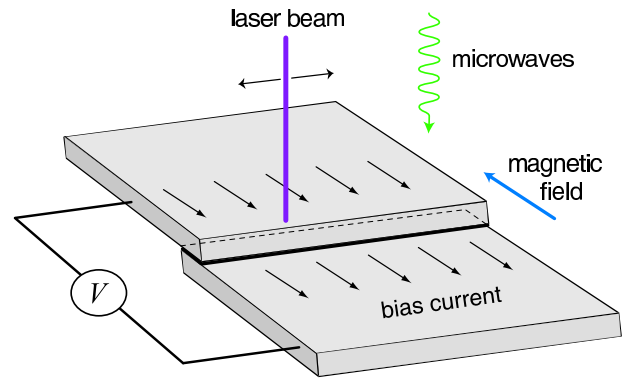}
\caption{Schematic view of the laser scanning experiment.
Dimensions are not to scale. \label{scheme}}
\end{center}
\end{figure}

In order to experimentally obtain information about the
microwave-induced vortex motion in the junction we have performed
spatially-resolved measurements using a low-temperature Laser
Scanning Microscope (LSM).

The principle of probing the junction by laser beam is to
introduce a local heating of the sample at the beam position and
to get the information about the local properties of the junction
at this point. We scanned a focused laser beam over the junction
area and simultaneously recorded its electrical response. The
voltage response of a current-biased junction is due to the change
of damping induced by local temperature increase at the
illuminated spot. In order to improve the signal-to noise ratio
and increase spatial resolution, the beam intensity is modulated
at a frequency $f_{\rm mod}$ and the circuit response is detected
using lock-in technique. The spatial resolution of this technique
is limited by both the thermal healing length (depending on the
material properties and $f_{\rm mod}$) and the laser beam
diameter. In the presented experiment it was about $2\,\mu {\rm
m}$.

Figure~\ref{scheme} illustrates the scheme of the LSM experiment.
The junction area is scanned by tightly-focused laser beam, which
power is modulated at $f_{\rm mod}=100\, {\rm kHz}$. The voltage
response $\delta V$ arising due to the temperature oscillations is
detected by lock-in technique. The characteristic time of a laser
beam displacement is much longer than $1/f_{\rm mod}$, which
allows to measure the voltage response for every geometrical point
of the junction averaged  over many thermal cycles. In its turn,
the characteristic modulation time $1/f_{\rm mod}$ is much longer
than the flux-flow oscillation frequency $V/\Phi_0$ in the
junction, where $\Phi_0=h/(2e)$ is the magnetic flux quantum. We
adjusted the laser beam power to have the relative magnitude of
the voltage signal $|\Delta V/V|\simeq 10^{-3}$, which ensured
that the beam acts as a weak perturbation of the dynamical state
of the junction. In order to reduce noise, we averaged the
measured $\delta V$ over the junction width. We plot the inverted
voltage response $(-\delta V)$ as a function of the laser beam
position along the length of the junction.

\begin{figure}[hbt]
\includegraphics[width=3.2in]{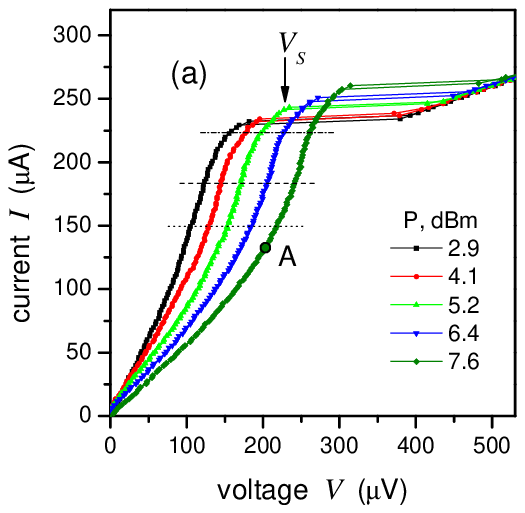}
\includegraphics[width=3.2in]{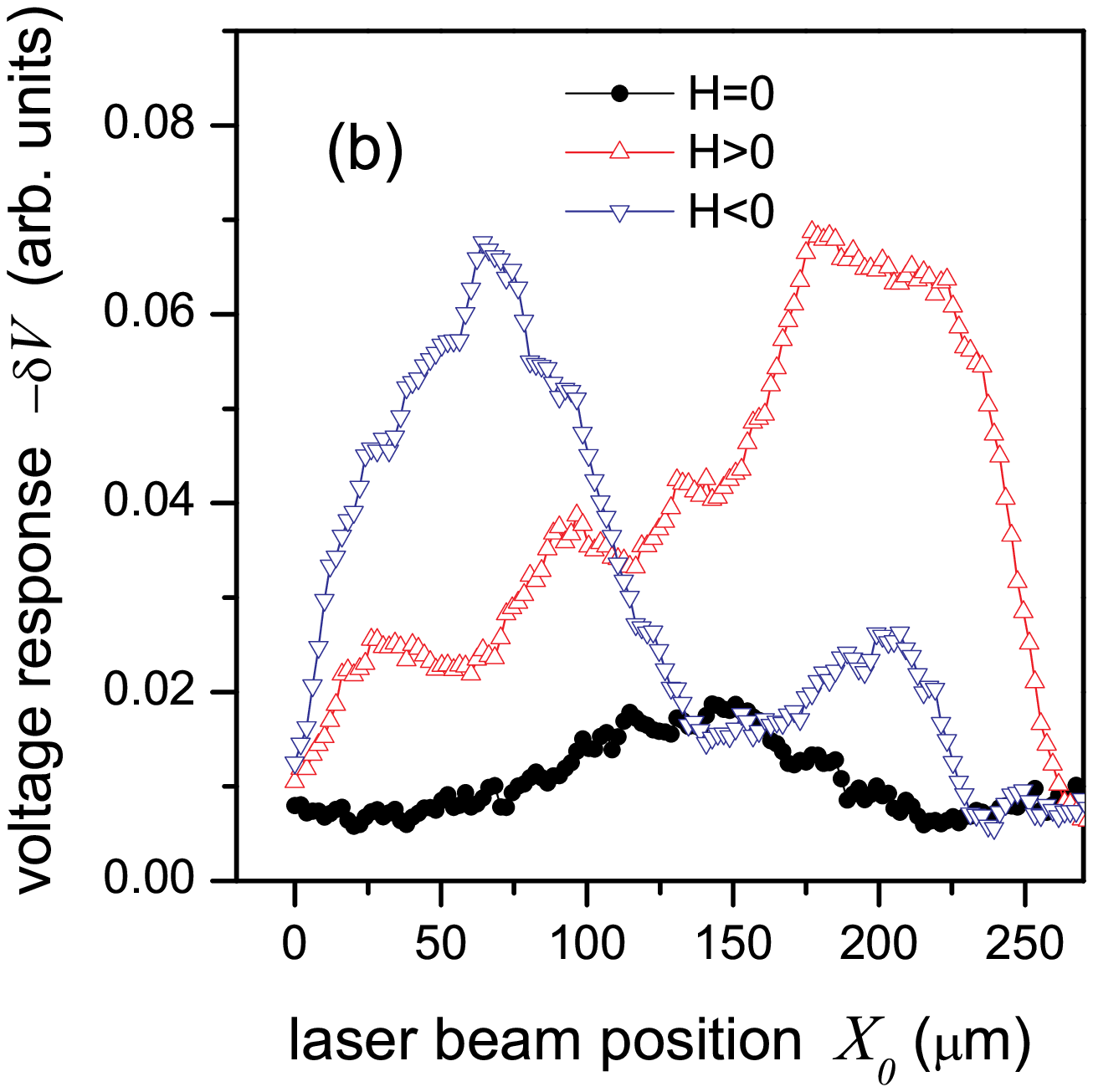}
\caption{(a) $I$-$V$ curves for different power levels
obtained at the same regime as ones from Fig.~\ref{Fig1}(b) but at
higher temperature $T=5\,$K; (b) LSM response of the junction in a
state corresponding to the marked bias point A in (a), at zero
(solid circles) and non-zero dc magnetic fields ($H=2.3\,$Oe up
triangles, $H=-2.3\,$Oe down triangles). } \label{Fig2}
\end{figure}

Results of the LSM investigation of the junction state
corresponding to point A marked on the $I$-$V$ curve in
Fig.~\ref{Fig2}(a) are presented in Fig.~\ref{Fig2}(b). The
vertical axis displays the inverted voltage signal $(-\delta V)$
in arbitrary units. Solid circles illustrate the LSM response in
the absence of dc external magnetic field. There is hardly any
voltage response at the junction edges, while a noticeable maximum
occurs in the center of the junction. The pattern is rather
symmetric reflecting the symmetry of the junction at zero magnetic
field.

To check the above hypothesis of the flux-flow nature of the
observed microwave driven behavior of the junction, we also
measured LSM patterns at non-zero dc magnetic field directed
perpendicular to both current flow direction and junction length
as shown in Fig.~\ref{scheme}. The main idea of applying dc field
is to break the symmetry between two junction edges and facilitate
injection of Josephson vortices at one of them. We would expect
that under such conditions the vortices would be mainly entering
the junction from one edge. The measured response patterns
corresponding to two magnetic fields of 2.3 Oe equal in magnitude
but opposite in polarity are presented in Fig.~\ref{Fig2}(b) by
up- and down-triangles. We clearly observe asymmetric patterns:
For $H>0$ a maximum of the response is located at the right half
of the junction while in the case of $H<0$ the maximum is shifted
to the opposite junction edge. We obtained similar asymmetric
patterns in non-zero magnetic field for all measured bias points
of $I$-$V$ curves in the Fig.~\ref{Fig2}(a), as well as for bias
points in Fig.~\ref{Fig1}(b). The maximum in the voltage response
$-\delta V$ was shifting back and forth between junction edges,
depending on the polarity of the applied field. This fact
indicates that the response is most probably generated by
Josephson vortices entering the junction from one of the edges.


We note that similar asymmetric response patterns were observed
for a pure flux-flow state in a long Josephson junction by Quenter
et al.~\cite{Quenter} and analyzed in Ref.~\cite{FistUst-LTSM}.
The response is typically the largest in the region where vortices
are injected in the junction. In detail, the effect of a local
perturbation induced by a hot spot (e.g., laser beam) on vortex
dynamics in the junction has been studied elsewhere
\cite{Malomed-93}.

Thus, we argue that junction dynamics under the microwave
irradiation at relatively low frequency $f\ll f_p$ is governed by
the injection of Josephson vortices at the junction edges. This
hypothesis is further supported by numerical simulations presented
in the following section. \label{Sec:SimRes}

\section{Numerical results}


The ac-driven weakly damped long Josephson junction is described
by the well-known perturbed sine-Gordon equation for the Josephson
phase difference $\varphi(x,t)$
\begin{equation}
\varphi _{xx}-\varphi _{tt}-\sin \varphi =\alpha(x) \varphi
_{t}-\gamma- \gamma_{\rm ac}\sin (\omega t)\:,  \label{Eq:sG}
\end{equation}
where the subscripts stand for the partial derivatives, the
coordinate $x$, running along the junction, and the time $t$ are
normalized so that the Josephson penetration length $\lambda_{\rm
J}$ and plasma frequency $\omega_{\rm p}=2\pi f_{\rm p}$ are both
equal to unity. A parameter $\alpha(x)$ is the damping
coefficient, which we will assume to be spatially dependent when
simulating the effect of the laser beam on the junction. The term
$\gamma$ is the uniform bias current density normalized to the
critical current density $j_c$, and $\gamma _{\rm ac}$ is the
amplitude of uniformly-flowing ac current normalized in the same
way.

If the long junction is placed in external dc and ac magnetic
field, Eq. (\ref{Eq:sG}) is accompanied by the boundary
conditions:
\begin{eqnarray}
\varphi_x(0,t) &=&h-h_{\rm ac}\sin (\omega t),  \label{Eq:BC1} \\
\varphi_x(\ell,t) &=&h+h_{\rm ac}\sin (\omega t), \label{Eq:BC2}
\end{eqnarray}
where $\ell=L/\lambda_{\rm J}$ is the normalized length of the
junction. In (\ref{Eq:BC1}) and (\ref{Eq:BC2}) we take opposite
signs in front of $h_{\rm ac}$ assuming that a part of ac bias
(due to its nonuniform distribution) flows near the junction edges
\cite{Olsen-Samuelsen:83,Samuelsen-Vasenko:85}. The precise
distribution of the oscillatory currents and fields induced by
microwaves in the junction can be rather complex. We suppose that
in the absence of a ground plane the largest microwave current is
induced near the junction edges. That can be described by a
distribution of the current $\gamma _{\rm ac}$ sharply peaked near
the junction boundaries \cite{Goldobin-2002} or, equivalently, by
an oscillating bias component $\pm\,h_{\rm ac}$ at the edges. In
the following we show the simulation results obtained for
$\gamma_{\rm ac}=0$, so that microwave drive is coupled via the
boundary conditions. We have also done simulations with a
uniformly distributed $\gamma _{\rm ac}$ current and got
qualitatively similar results.

We numerically integrated Eq.~(\ref{Eq:sG}) with boundary
conditions (\ref{Eq:BC1}) and (\ref{Eq:BC2}). The details of the
simulation procedure can be found elsewhere \cite{Malomed-93}.
Simulations were performed with parameters very close to our
experiment described above: a normalized junction length
$\ell=8.9$, the normalized ac-drive frequency $\omega/(2\pi)=0.2$,
and two values of dissipation coefficient $\alpha(x)=\alpha_0 =
0.05$ and $\alpha_0 =0.02$. The calculated averaged voltage
$\Theta$ is just $\langle\dot{\varphi}\rangle$ normalized to the
voltage at the first zero-field step (ZFS1)
$\langle\dot{\varphi}\rangle_{\rm ZFS1}=2\pi/\ell$.

\begin{figure}[b]
\includegraphics[width=3.2in]{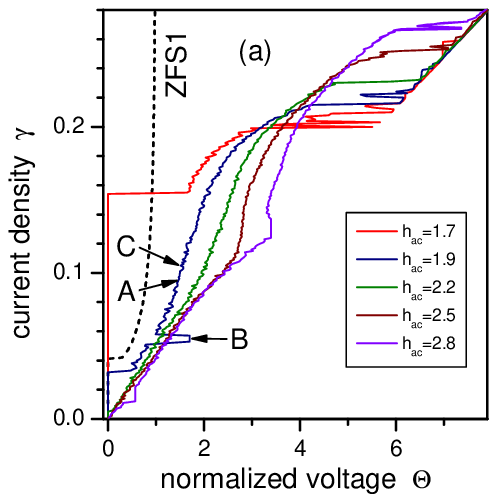}
\end{figure}
\begin{figure}
\includegraphics[width=3.2in]{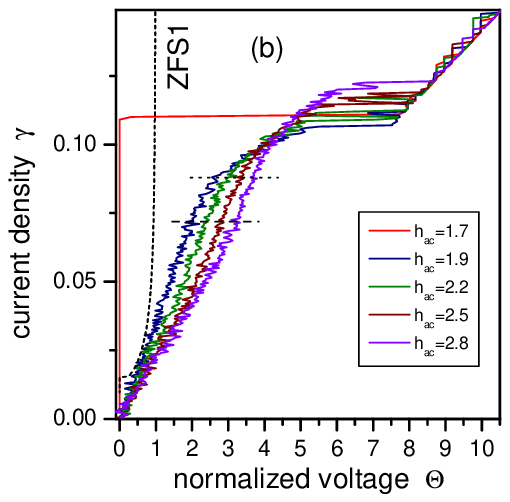}
\includegraphics[width=3.2in]{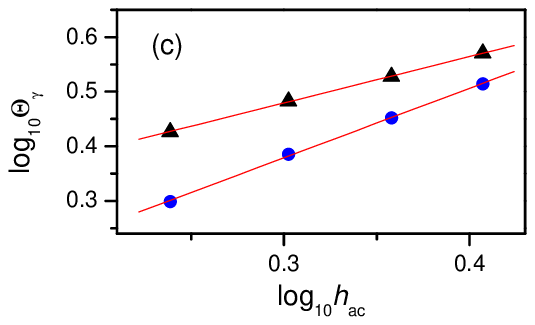}
\caption{Numerically calculated $I$-$V$ curves for
different amplitudes of the ac bias and two different values of
dissipation: (a) $\alpha_0 = 0.05$; (b) $\alpha_0 =0.02$. Dashed
IV-curve shows the first zero-field step (ZFS1) corresponding to a
vortex shuttling between junction boundaries in the absence of
microwaves. (c) Dependence of the voltage $\Theta_{\gamma}$ at the
dc bias levels $\gamma=0.072$ (circles) and $\gamma=0.088$
(triangles) indicated by horizontal dotted lines in (b) on
ac-drive amplitude $h_{\rm ac}$ using double-logarithmic scale.}
\label{I-V-numerics}
\end{figure}

Figure~\ref{I-V-numerics}(a,b) shows the numerically calculated
current-voltage characteristic for several values of the
normalized ac field amplitude $h_{\rm ac}$ at zero dc magnetic
field $h=0$. As the amplitude of ac drive $h_{\rm ac}$ approaches
the amplitude $h_{\rm ac}=2$, the critical current is reduced and
the $I$-$V$ curves display a smooth branch at low voltages. In
this regime, in order to obtain accurate enough curves we had to
perform numerical integration over 40,000 time units at every bias
point. Increasing the amplitude $h_{\rm ac}$ shifts the smooth
branch towards high voltages. In some current ranges we observe
locking of the $I$-$V$ curves to vertical Shapiro steps at
constant voltage. Figures~\ref{I-V-numerics}(a) and
\ref{I-V-numerics}(b) correspond to two different damping
constants $\alpha_0 = 0.05$ and $\alpha_0 =0.02$, respectively.

The dependence of the voltage $\Theta_{\gamma}$ taken at the
constant dc bias levels $\gamma=0.072$ and $\gamma=0.088$ on the
applied ac-drive amplitude $h_{\rm ac}$ is presented in
Fig.~\ref{I-V-numerics}(c). The best linear fits to the power law
$\Theta_I\propto h_{\rm ac} ^{\beta}$ are with the indices
$\beta=1.27\pm 0.08$ and $\beta=0.85\pm 0.08$ for smaller and
larger current values, correspondingly.

Thus, we find in good accord with the experimental results that
the characteristic voltage values of the low voltage branch depend
on the microwave power as $V~\propto~ P^{1/2}$ (see Fig.
\ref{I-V-numerics}(c)). By analogy with the flux-flow behavior, we
argue that the characteristic voltage $V$ is proportional to the
average number of fluxons in the junction, which in its turn is
roughly proportional to the microwave field intensity $\sim
P^{1/2}$. We should take here into account that the microwave
current generates magnetic field at the junction boundaries. At
every time moment, the generated magnetic field components at the
two long junction boundaries are equal in magnitude but opposite
in polarity. If the field amplitude is larger than critical, there
will be vortices injected at one junction boundary and
antivortices injected at the other boundary. At microwave
frequencies $f\ll f_{\rm p}$ the magnetic field generated at the
junction boundaries varies relatively slow in comparison with the
vortex injection rate. Thus, at each junction boundary, there will
be a bunch of vortices injected every half-period of microwave
field and a bunch of antivortices injected at every other
half-period. During half-period of the microwave oscillation,
vortices on one side of the junction and antivortices on its other
side are injected simultaneously and propagate through the
junction under the influence of the dc bias current. Vortices and
antivortices will eventually annihilate in the middle of the
junction or, as the dissipation is rather low, may propagate
through one another and annihilate at the opposite junction
boundary.

\begin{figure}[htb]
\includegraphics[width=3.2in]{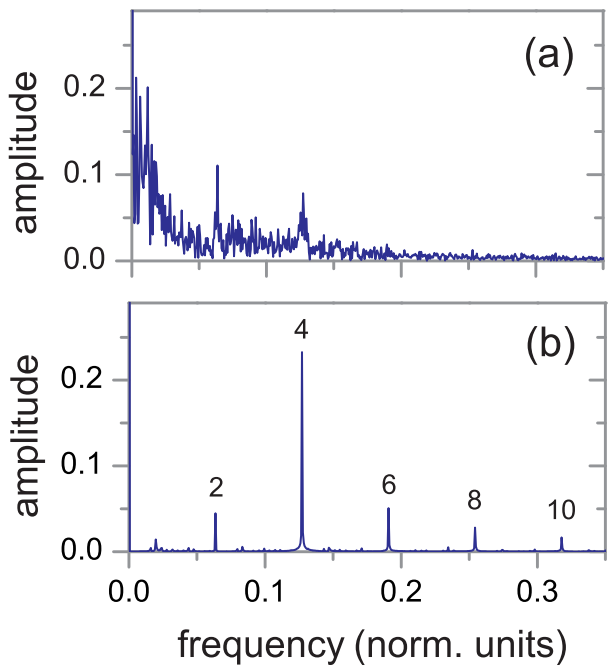}
\caption{Calculated voltage spectra at two bias points of the
current-voltage curves shown in Fig.~\ref{I-V-numerics}: (a) point
A; (b) point B. The numbers near the peaks indicate the
corresponding harmonics of the ac drive frequency.}
\label{spec-numerics}
\end{figure}

We found that voltage oscillations in the junction are very {\it
irregular} at nearly any point of smooth $I$-$V$ branches. We
calculated the Fourier power spectra of the voltage oscillations
$\dot\varphi(t)$ at the center of the junction using FFT.
Figure~\ref{spec-numerics} illustrates such spectra for two points
A and B of the current-voltage characteristics shown in
Fig.~\ref{I-V-numerics}(a). Point A displays a typical spectrum
for the states along the smooth branches (see
Figure~\ref{spec-numerics}(a)). Voltage oscillations are strongly
irregular, the oscillation spectrum is very broad and resembles a
chaotic state. Point B has been chosen at a Shapiro step, which is
expected to be synchronized with the frequency of ac drive
$f=\omega/(2\pi)\approx 0.0318$. In the corresponding spectrum
presented in Fig.~\ref{spec-numerics}(b) we clearly see strong
peaks at frequencies corresponding to various harmonics of the
driving frequency. We suppose that the prevailing even harmonics
are related to symmetric injection of vortices and antivortices
into the junction from two boundaries.

Next we turn to a numerical simulation of our LSM experiment where
the laser beam produces a local heating of the junction around the
beam position at $x=x_0$.  Thereby the dissipative term $\alpha$
is locally increased. We model the laser beam perturbation by a
dissipative spot placed at the point $x=x_0$ and described by the
following function:
\begin{equation}
\alpha(x) = \alpha_0 \left[1 + \varepsilon\delta(x - x_0)\right]
\:.
\end{equation}
In simulations, we approximate the $\delta-$function by the
expression \cite{Malomed-93}
\begin{equation}
\varepsilon\delta (x-x_0) \approx
 \kappa\left[ 1 - \tanh^{2} \frac{ 2
(x-x_0)}{\eta}\right]\:,
\end{equation}
where $\varepsilon=\kappa\eta$. In order to model the laser beam
scanning procedure we calculate the time averaged voltage across
the Josephson junction as a function of $x_0$ . This algorithm
accounts for the experimental situation since the period of the
fluxon oscillations is much shorter than the characteristic times
of scanning and power modulation of the laser beam. Similarly to
the experiment, the LSM patterns are presented in form of the
dependence of the inverted voltage change $-\Delta \Theta (x_0)$
at a fixed bias point on the $I$-$V$ curve. In simulations we have
chosen the strength of the laser spot given by $\kappa=2$
$\eta=1$.

\begin{figure}[htb]
\includegraphics[width=3.2in]{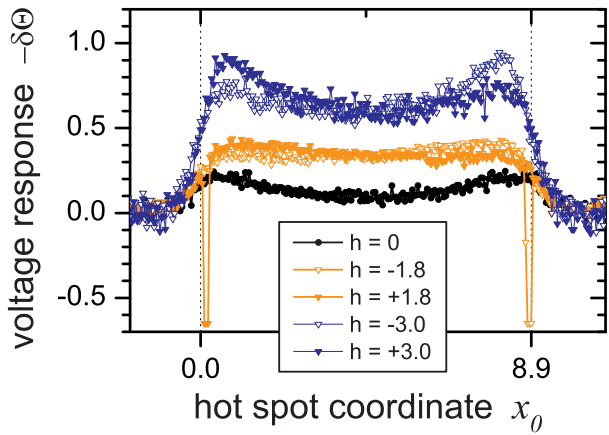}
\caption{Simulated LSM response of the junction in a state
corresponding to the marked bias point C in
Fig.~\ref{I-V-numerics}(a), at zero (solid circles) and non-zero dc
magnetic field (solid triangles for positive fields and open
triangles for negative fields).} \label{scan-numerics}
\end{figure}

Figure \ref{scan-numerics} shows the calculated voltage response
$-\Delta \Theta(x_0)$ obtained for the bias point C marked in
Fig.~\ref{I-V-numerics}. The scan shown by solid circles has been
simulated at $h=0$. Two types of triangles show simulation results
for positive and negative dc magnetic field. These plots should be
compared to experimentally measured ones presented in
Fig.~\ref{Fig2}(b). One can see that simulations qualitatively
match the experiment. In both cases, the response increases with
the magnitude of magnetic field and becomes asymmetric. Inverting
the field reverses the response with respect to the center of the
junction. The magnetic field makes it favorable for vortices (or
antivortices for negative fields) to enter from one boundary of
the junction. The response is the largest near the boundary
\cite{Quenter,FistUst-LTSM}, i.e. in the region where vortices
enter the junction.

\section{Discussion and Conclusions}
\label{Sec:Conclusion}

Our numerical studies show that applying of a microwave radiation
of large amplitude induces a vortex flow in long Josephson
junctions.  As the frequency of ac drive is small, i. e. $f \ll
f_p$, the dynamics of vortices is extremely irregular. It is
appropriate to note here that chaotic dynamics in long junctions
driven by ac field has been observed and discussed before
\cite{chaos-LJJ-with-RF}. However, the novelty of our results is
in the particular appearance of the chaotic states in $I$-$V$
curves, which we relate to experimental observations. Indeed, the
observed low-voltage branches are similar to the well-known
displaced linear branch, which is associated with chaotic dynamics
in the absence of microwaves \cite{chaos-LJJ-no-RF}. In both
cases, the smooth branches shift towards high voltages with
increasing the amplitude of the field (either ac or dc). This
feature becomes even more pronounced, displaying the downward
curvature (see Figs. \ref{Fig2}(a) and \ref{I-V-numerics}(a)) as
the chaos in vortex dynamics is reduced. Note here that the
increase of temperature (and thus increase of dissipation)
suppresses the chaos.

\begin{figure}[htb]
\includegraphics[width=2.8in]{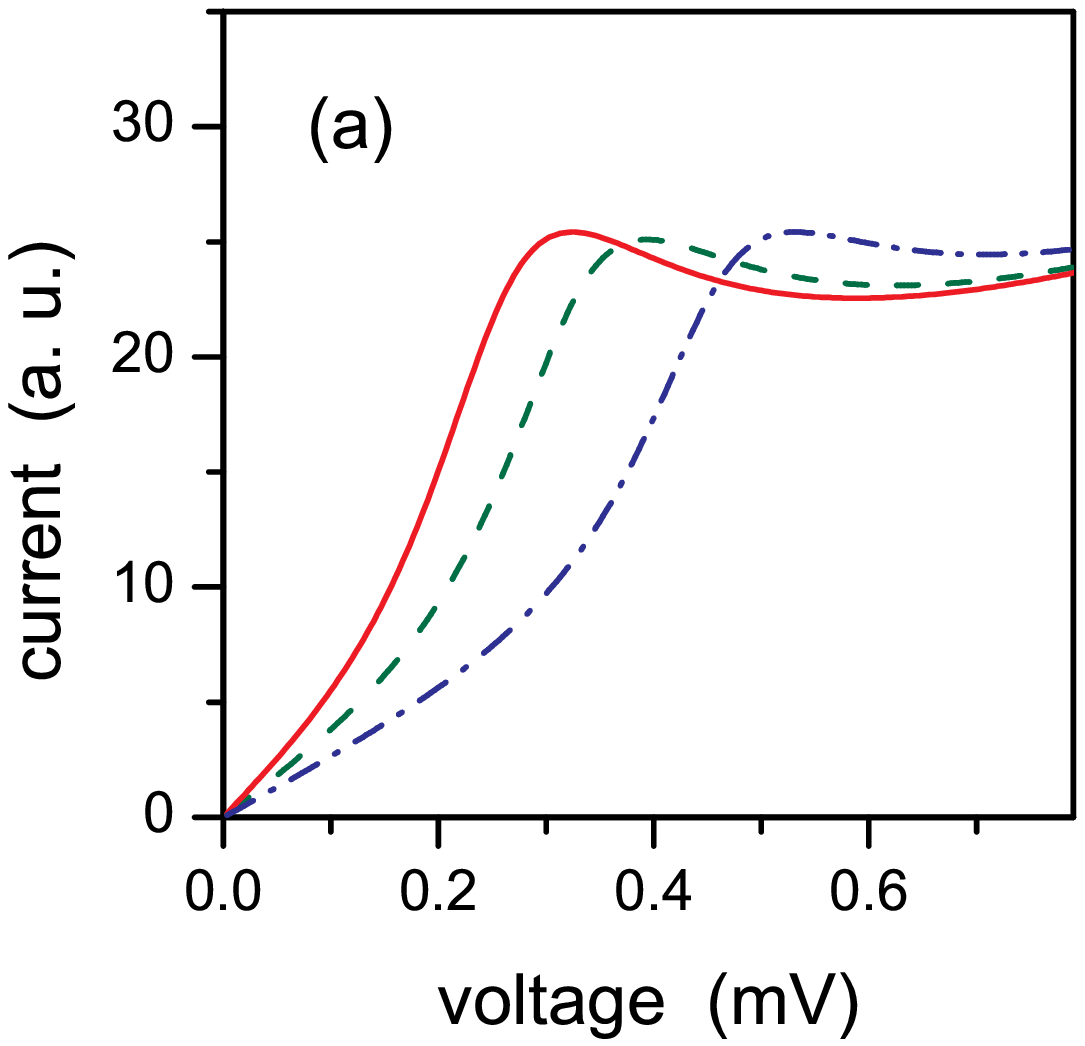}
\includegraphics[width=2.8in]{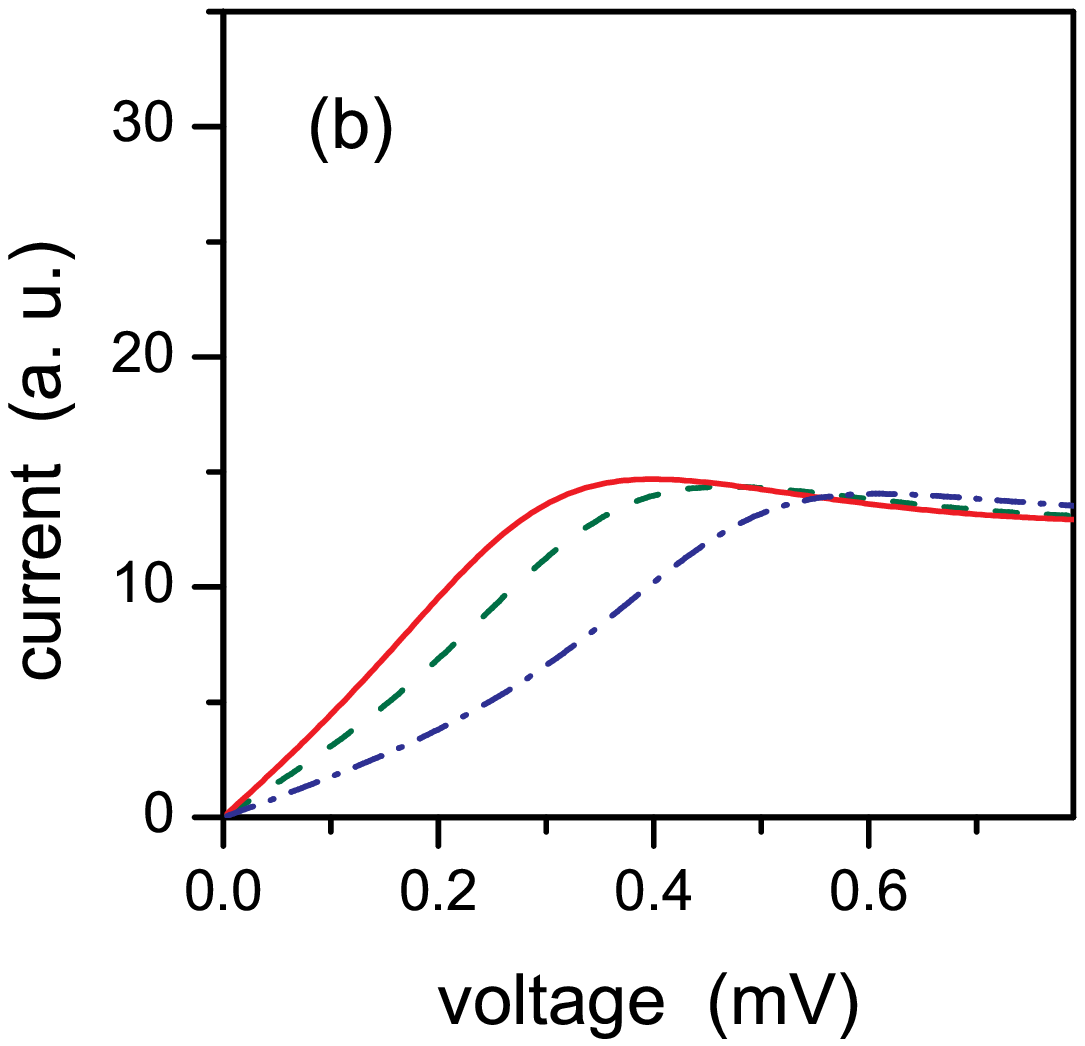}
\caption{The power dependence of $I$-$V$ curves calculated using
Eq. (\ref{SCanalitik}) for two values of parameter $\delta$: (a)
$\delta=0.0001$; (b) $\delta=0.0002$. The power levels correspond
to $P=2, ~5,~ 7$ dBm. In both cases the linear slope corresponding
to quasiparticle resistance $R=50$ k$\Omega$ (a) and $R=140$
k$\Omega$ (b), is added. }
\end{figure}

The presence of strong irregularities in the vortex motion allows
us to draw an analogy with the properties of extremely small
Josephson junctions subject to microwave radiation in the presence
of large thermal fluctuations \cite{KovalFistUSt}. Indeed, in this
case smooth low voltage branches in the $I$-$V$ curve have been
observed, and these branches also shift towards larger voltages
proportionally to $\sqrt{P}$. This effect has been explained as an
incoherent multi-photon absorption by the Josephson phase
described by the diffusive motion induced by thermal noise. In
this case we can write the lumped Josephson phase as
\begin{equation} \label{JosphaseMicrowave}
\varphi(t)~=~Vt+\psi(t)-\frac{\eta}{\alpha}\cos\omega t +
\varphi_1(t)~~,
\end{equation}
where $\eta$ is the amplitude of ac drive, the random function
$\psi(t)$ determines the Josephson phase diffusion, and
$\varphi_1(t)$ has to be found self-consistently
\cite{KovalFistUSt}. As the frequency $\omega$ is small, we have
obtained the low voltage branch of $I$-$V$ curve in the following
analytical form:
\begin{equation} \label{SCanalitik}
I_s(V)=I_c\frac{1}{2\alpha} [f_+(V-V_s)f_+(V+V_s)-f_-(V-V_s)f_-(V+V_s)]~~,
\end{equation}
where the functions $f_{\pm}(x)~=~\frac{ \sqrt{
\sqrt{x^2+\delta^2} \pm x }} { \sqrt{x^2+\delta^2} } $. Here, the
maximum voltage of the branch
$V_s~=~\eta/\alpha~\propto~\sqrt{P}$, and $\delta$ is the
parameter characterizing the strength of fluctuations. The $I$-$V$
curves are presented in Fig. 7 for different values of microwave
power and two values of $\delta$. Note here, that the low voltage
branch appears in the limit of $\omega \ll \delta$ as the thermal
fluctuations wash out the Shapiro steps. As the fluctuations
characterized by the parameter $\delta$ increase, the $I$-$V$
curve gets more linear (see Fig. 7(b)).

We see that all features obtained for microwave induced irregular
vortex flow in long junctions resemble ones observed in a small
microwave driven Josephson junction in the presence of thermal
fluctuations. Thus we argue that the ansatz
(\ref{JosphaseMicrowave}) can be qualitatively used in the case of
a microwave induced vortex flow, where the random function
$\psi(t)$ is due to chaos in vortex motion. The chaos in the
vortex motion mimics thermal fluctuations and wash out the $I$-$V$
curve. According to both our experiment and numerical study, the
reduction of chaos by increasing the damping (i. e. decreasing
$\delta$) leads to the pseudo-resonant feature on the $I$-$V$
curve (see Figs. \ref{Fig2}(a) and \ref{I-V-numerics}(a)). It is
interesting to note that an increase of temperature (dissipation)
in the case of long Josephson junction leads to \emph{smaller}
chaos-induced fluctuations. This is opposite to the case of a
small Josephson junctions where the parameter $\delta$ increases
with an increase of temperature.

In summary, we reported here experimental and numerical study of
microwave-induced effects on the current-voltage characteristics
of long Josephson junctions. We observed that low-frequency
microwaves act on the junction $I$-$V$ curves similar to the
effect of dc magnetic field, i.e. induce the flux-flow behavior.
Experiments using a low-temperature laser scanning microscope
unambiguously indicate the motion of Josephson vortices driven by
microwaves. Numerical simulations are in a good agreement with
experimental data and indicate that the vortex flow is strongly
irregular. Our results provide also an explanation for previously
measured $I$-$V$ curves of intrinsic high-$T_{\rm c}$ Josephson
junctions driven by microwaves \cite{HTS-rf-branches}.

When this work was completed, we also learned about recent
numerical data of the Tuebingen group \cite{Tuebingen-2004}. Their
simulations using the stacked junction model with the parameters
of intrinsic junctions also show chaotic regimes in the vortex
dynamics driven by microwaves, which is consistent with our
observations.

\section*{Acknowledgements}

We are indebted to R. Kleiner and H.B. Wang for valuable
discussions.


\begin{thebibliography}{}

\bibitem{kleiner92}
R. Kleiner, F. Steinmeyer, G. Kunkel, and P. M{\"u}ller, Phys.
Rev. Lett. {\bf  68},  2394   (1992); R. Kleiner and P. M\"uller,
{\it Phys. Rev. B} {\bf 49}, 1327 (1994).

\bibitem{plasma-res-HTS} O.K.C. Tsui, N.P. Ong, Y. Matsuda,
Y.F. Yan, and J.B. Peterson, Phys. Rev. Lett. 73, 724 (1994); Y.
Matsuda, M.B. Gaifullin, K. Kumagai, K. Kadowaki and T. Mochiku,
Phys. Rev. Lett. \textbf{75}, 4512 (1995); L.N. Bulaevskii; M.P.
Maley, and M. Tachiki, Phys. Rev. Lett. \textbf{74}, 801 (1995);
A.E. Koshelev, Phys. Rev. Lett. \textbf{77}, 3901 (1996).

\bibitem{HTS-rf-branches} A. Irie and G. Oya, IEEE Trans. Appl.
Supercond. \textbf{5}, 3267 (1995); Yu.I. Latyshev, P. Monceau,
and V.N. Pavlenko, Physica C \textbf{293}, 174 (1997); W.
Prusseit, M. Rapp, K. Hirata, and T. Mochiku, Physica C
\textbf{293}, 25 (1997); H.B. Wang, Y. Aruga, T. Tachiki, Y.
Mizugaki, J. Chen, K. Nakajima, T. Yamashita, and P.H. Wu, Appl.
Phys. Lett. \textbf{74}, 3693 (1999); Y.J. Doh, J.H. Kim, H.S.
Chang, S.H. Chang, H.J. Lee, K.T. Kim, W. Lee, and J.H. Choy,
Phys. Rev. B \textbf{63}, 144523 (2001).

\bibitem{Costabile} G. Costabile, R. Monaco, S. Pagano, and G.
Rotoli, Phys. Rev. B \textbf{42}, 2651 (1990).

\bibitem{locking} M.~Salerno, M.~R.~Samuelsen,
G.~Filatrella, S.~Pagano, and R.~D.~Parmentier, Phys. Rev. B {\bf
41}, 6641 (1990); N.~F.~Pedersen and A.~Davidson, Phys. Rev. B
{\bf 41}, 178 (1990); N.~Gr{\o}nbech-Jensen, Phys. Rev. B {\bf
47}, 5504 (1993); N.~Gr{\o}nbech-Jensen and M.~ Cirillo, Phys.
Rev. B {\bf 50}, 12851 (1994); M.~Cirillo, P.~Cocciolo, V.~Merlo,
N.~Gr{\o}nbech-Jensen, and R.~D.~Parmentier, J. Appl. Phys. {\bf
75}, 2125 (1994).

\bibitem{chaos-LJJ-with-RF} M. Cirillo, J. Appl. Phys. \textbf{60},
338 (1986); M. Cirillo, A.R. Bishop, N. Gr{\o}nbech-Jensen and
P.S. Lomdahl, Phys. Rev. E \textbf{49}, 3606 (1994).

\bibitem{chaos-LJJ-no-RF} A. V. Ustinov, H. Kohlstedt, and P.
Henne, Phys. Rev. Lett. \textbf{77}, 3617 (1997); M. Cirillo, T.
Doderer, S. G. Lachenmann, F. Santucci, and N. Gr{\o}nbech-Jensen,
Phys. Rev. B \textbf{56}, 11889 (1997).

\bibitem{Goldobin-2002} E. Goldobin, A.M. Klushin, M. Siegel,
and N. Klein, J. Appl. Phys. {\bf 92}, 3239 (2002).

\bibitem{Kosh-Shitov-2000} V.P. Koshelets and S.V. Shitov,
Supercond. Sci. Technol. \textbf{13}, R53 (2000).

\bibitem{KovalFistUSt} Y. Koval, M. V. Fistul and A. V. Ustinov,
submitted to Phys. Rev. Lett. (2004).

\bibitem{Jena} The sample was fabricated using our layout
at the foundry of IPHT Jena, Germany
(\textsl{http://www.ipht-jena.de}).

\bibitem{Leon-2002} F. Geniet and J. Leon, Phys. Rev. Lett.
{\bf 89}, 134102 (2002).

\bibitem{Borya-Misha} M. V. Fistul, B. A. Malomed, and A. V. Ustinov
(unpublished).

\bibitem{plasma-LJJ} M. Cirillo, G. Costabile, S. Pace, and B.
Savo, IEEE Trans. Magn. \textbf{19}, 1014 (1983); T. Holst and J.
Bindslev Hansen, Phys. Rev. B \textbf{44}, 2238 (1991); M.V.
Fistul and A.V. Ustinov, Phys. Rev. B \textbf{63}, 024508 (2001);

\bibitem{Pedersen-Sakai} N.F. Pedersen and S. Sakai, Phys. Rev.
B \textbf{58}, 2820 (1998); S. Sakai and N.F. Pedersen, Phys. Rev.
B \textbf{60}, 9810 (1999).

\bibitem{Quenter} D. Quenter, A. V. Ustinov, S. G. Lachenmann,
T. Doderer, R. P. Huebener, F. M{\"u}ller, J. Niemeyer, R.
P{\"o}pel, and T. Weimann, Phys. Rev. B \textbf{51}, 6542 (1995).

\bibitem{FistUst-LTSM} M. V. Fistul and A. V. Ustinov,
Inst. Phys. Conf. Ser, {\bf 167}, 177 (2000).

\bibitem{Olsen-Samuelsen:83} O. H. Olsen and M. R. Samuelsen,
J. Appl. Phys. \textbf{54}, 6522 (1983).

\bibitem{Samuelsen-Vasenko:85} M. R. Samuelsen and S. A. Vasenko,
J. Appl. Phys. \textbf{57}, 110 (1985).

\bibitem{Malomed-93} B. A. Malomed and A. V. Ustinov, Phys. Rev.
B \textbf{49}, 13024 (1994).

\bibitem{Tuebingen-2004} T. Clauss, T. Uchida, M. M\"o{\ss}le,
D. Koelle, and R. Kleiner (unpublished).

\end{thebibliography}
\end{document}